\documentclass[prd,aps,onecolumn,tightenlines,notitlepage,nofootinbib,preprintnumbers,letterpaper,superscriptaddress]{revtex4-1}  

\pdfoutput=1
\usepackage{epsfig}
\usepackage{amsfonts}
\usepackage{amsmath}
\usepackage{graphicx}
\usepackage{color}
\usepackage{xcolor}
\usepackage{amssymb,amsmath,hyperref,slashed,color,graphicx,bm, url}
\usepackage[T1]{fontenc}
\usepackage{relsize}
\usepackage{marvosym}
\usepackage{booktabs} 
\usepackage{accents}
\usepackage{mathrsfs}

\DeclareMathAlphabet{\mathpzc}{OT1}{pzc}{m}{it}
\DeclareMathAlphabet{\mathbf}{U}{bf}{m}{n}
\DeclareMathAlphabet{\mathfrak}{U}{frak}{m}{n}

\newcommand{\keV}{{\rm keV}}

\graphicspath{{Figures/}}

\begin{document}

\hfill{FERMILAB-PUB-20-483-T, NUHEP-TH/20-09, IFT-UAM/CSIC-20-131, FTUAM-20-19}
\title{The physics potential of a reactor neutrino experiment with Skipper~CCDs: Measuring the weak mixing angle}

\author{Guillermo Fernandez-Moroni}
\email[E-mail:]{gfmoroni@fnal.gov}
\thanks{\scriptsize \!\! \href{https://orcid.org/0000-0002-1654-9562}{0000-0002-1654-9562}}
\affiliation{Fermi National Accelerator Laboratory, Batavia, IL, 60510, USA}

\author{Pedro~A.~N.~Machado}   
\email[E-mail:]{pmachado@fnal.gov}
\thanks{\scriptsize \!\! \href{https://orcid.org/0000-0002-9118-7354}{0000-0002-9118-7354}}
\affiliation{Fermi National Accelerator Laboratory, Batavia, IL, 60510, USA}

\author{Ivan Martinez-Soler}   
\email[E-mail:]{ivan.martinezsoler@northwestern.edu}
\thanks{\scriptsize \!\! \href{https://orcid.org/0000-0002-0308-3003}{0000-0002-0308-3003}}
\affiliation{Fermi National Accelerator Laboratory, Batavia, IL, 60510, USA}
\affiliation{Northwestern University, Department of Physics \& Astronomy, 2145 Sheridan Road, Evanston, IL 60208, USA}
\affiliation{Colegio de F\'isica Fundamental e Interdisciplinaria de las Am\'ericas (COFI), 254 Norzagaray street, San Juan, Puerto Rico 00901.}

\author{Yuber~F.~Perez-Gonzalez}   
\email[E-mail:]{yfperezg@northwestern.edu}
\thanks{\scriptsize \!\! \href{https://orcid.org/0000-0002-2020-7223}{0000-0002-2020-7223}}
\affiliation{Fermi National Accelerator Laboratory, Batavia, IL, 60510, USA}
\affiliation{Northwestern University, Department of Physics \& Astronomy, 2145 Sheridan Road, Evanston, IL 60208, USA}
\affiliation{Colegio de F\'isica Fundamental e Interdisciplinaria de las Am\'ericas (COFI), 254 Norzagaray street, San Juan, Puerto Rico 00901.}

\author{Dario Rodrigues}
\email[E-mail:]{rodriguesfm@df.uba.ar}
\thanks{\scriptsize \!\! \href{https://orcid.org/0000-0002-7952-7168}{0000-0002-7952-7168}}
\affiliation{Fermi National Accelerator Laboratory, Batavia, IL, 60510, USA}
\affiliation{Department of Physics, FCEN, University of Buenos Aires and IFIBA, CONICET, Buenos Aires,
Argentina}

\author{Salvador Rosauro-Alcaraz}   
\email[E-mail:]{salvador.rosauro@uam.es}
\thanks{\scriptsize \!\! \href{https://orcid.org/0000-0003-4805-5169}{0000-0003-4805-5169}}
\affiliation{Departamento de F\'isica T\'eorica and Instituto de F\'isica T\'eorica, IFT-UAM/CSIC, Universidad Aut\'onoma
de Madrid, Cantoblanco, 28049, Madrid, Spain}

\begin{abstract}
    We analyze in detail the physics potential of an experiment like the one recently proposed by the vIOLETA collaboration: a kilogram-scale Skipper CCD detector deployed 12 meters away from a commercial nuclear reactor core. This experiment would be able to detect coherent elastic neutrino nucleus scattering from reactor neutrinos, capitalizing on the exceptionally low ionization energy threshold of Skipper CCDs. To estimate the physics reach, we elect the measurement of the weak mixing angle as a case study. We choose a realistic benchmark experimental setup and perform variations on this benchmark to understand the role of quenching factor and its systematic uncertainties, background rate and spectral shape, total exposure, and reactor antineutrino flux uncertainty. We take full advantage of the reactor flux measurement of the Daya Bay collaboration to perform a data driven analysis which is, up to a certain extent, independent of the theoretical uncertainties on the reactor antineutrino flux. We show that, under reasonable assumptions, this experimental setup may provide a competitive measurement of the weak mixing angle at few MeV scale with neutrino-nucleus scattering.
\end{abstract} 

\maketitle

\section{Introduction}
The recent discovery of coherent neutrino-nucleus scattering (CEvNS)~\cite{Freedman:1973yd} by the COHERENT experiment~\cite{Akimov:2017ade} has stirred a large deal of interest of the high energy physics community. 
Although experimentally challenging, the CEvNS cross section opens up a new, exciting venue to detect neutrinos.
The key reason lies in the fact that this is a relatively large neutrino scattering cross section. This is because when a neutrino transfers a small amount of momentum to a nucleus, with a de Broglie wavelength comparable to nuclear radii, the neutrino probes the nucleus as a whole, instead of scattering with individual nucleons. Therefore, the CEvNS cross section is sensitive to the square of the nucleus weak charge, ${\cal Q}_V^{\rm SM}\equiv N-(4s_W^2-1)Z$, where $s_W\equiv\sin\theta_W$ is the weak mixing angle and  $N,Z$ denote the number of neutrons and protons in the nucleus, respectively.
This is typically referred to as ``coherent enhancement'' and may enhance the cross section 100-fold or even more, compared to typical quasi-elastic processes. 
Such a large enhancement opens up the possibility of employing relatively small detectors and still being able to probe interesting and novel physics in the neutrino sector.

Nevertheless, observing low energy nuclear recoils is still a challenge overcome by only few detectors. 
Among those are high purity germanium detectors, as used for example in the CoGeNT~\cite{Aalseth:2010vx} and GEMMA~\cite{Beda:2013mta} experiments, and dual phase xenon time projection chambers, such as those used in XENON1T~\cite{Aprile:2012zx}.
A more recent and promising technology, and the main focus of the present paper, are  Skipper charge couple devices (Skipper CCDs).

Skipper CCDs are 2D pixelated detectors fabricated on high resistivity silicon \cite{Holland:2003}. Its small pixel size of 15 $\mu$m by 15 $\mu$m allows for good spatial resolution and event discrimination. Devices with several millions of pixels can be fabricated with high yield. The high resistive substrate permits an active depth of each pixel of 675 $\mu$m totaling up to approximately 8 grams of active silicon per device. In particular, the great advantage of the Skipper CCD is the high energy resolution thanks to its very low readout noise. It provides single-charge counting capability for any ionization packet in the active volume \cite{Tiffenberg2017, skipper_2012}. The energy resolution is then limited by the silicon absorption process to the different particles. Extensive measurements of the absorption processes have been studied in these sensors using X-rays~\cite{Rodrigues2020}. When they are cooled close to liquid nitrogen temperatures the leakage current production is very small \cite{SENSEI_2019} and ionizations of few electrons can be unmistakably observed. Pixels are read sequentially by a single or just a few amplifiers per sensor using a pixel readout time of approximately 1 millisecond (for single-carrier counting capability). The slow readout provides poor time resolution. The technology is mature enough to allow the integration of many of these sensors running in parallel in a single system. Several groups are already pushing for the integration of many sensors for dark matter experiments: SENSEI experiment (200 grams) \cite{SENSEI_2019}, DAMIC-M~(1kg)~\cite{DAMICM}, and OSCURA (10 kg) \cite{oscura_2020}. Besides, CCDs have already been recognized as neutrino detectors~\cite{Moroni:2014wia}. 

The Skipper CCD capability of detecting such low energy nuclear recoils allows these detectors to observe neutrinos emitted from the most intense, known artificial source: commercial nuclear reactors. 
In nuclear reactors, the fission processes that generate heat (which is afterwards used to generate electricity) always involve copious amounts of neutrino emission. 
Each of these fission processes emit an average of 6 electron antineutrino and 200~MeV of heat.
This roughly translates into $2\times 10^{20}$ neutrinos per second per GW$_{th}$~\cite{Ma:2012bm}. 
To put things in perspective, the Spallation Neutron Source (SNS), the neutrino source used by the COHERENT collaboration, has an intensity of about $5\times 10^{20}$ protons on target (POT) per day~\cite{Akimov:2017ade}. 
For SNS energies, each proton delivered on the target should lead to an average of about 0.24 neutrinos originated from the decay of pions produced in the beam.
Another example can be made with the future DUNE experiment which plans to deliver $1.1\times10^{21}$~POT per year of running, or yet a neutrino flux of $3.4\times10^6/{\rm cm}^2/{\rm s}$ in the near detector.
Despite the enormous neutrino flux, nuclear reactors emit neutrinos with much lower energies than the aforementioned sources, typically around the MeV scale.
A coherent scattering of those neutrinos would transfer not more than a couple hundred electronVolts to the recoiled nucleus.
This makes the use of such intense neutrino source very difficult to most detector technologies. 
Exceptions are the CONNIE~\cite{Aguilar-Arevalo:2016khx}, NUCLEUS~\cite{Angloher:2019flc} and CONUS~\cite{Buck:2020opf} experiments.
In addition to those, Skipper CCDs are in a special position to leverage nuclear reactors due to their very low energy thresholds.

The physics reach of several CEvNS experiments have been studied in the literature. 
For example, light mediators coupling to standard model particles have been searched for in the CONNIE experiment~\cite{Aguilar-Arevalo:2019zme}.
On the theoretical side, there are several studies evaluating the physics capabilities of CEvNS experiments, including the following:
CEvNS cross section measurements~\cite{Collar:2014lya, Skiba:2020msb}, 
nuclear physics probes~\cite{Coloma:2020nhf},
neutrino non-standard interactions~\cite{Coloma:2017ncl, Farzan:2017xzy, Liao:2017uzy, Esteban:2018ppq, AristizabalSierra:2018eqm, Altmannshofer:2018xyo, Abdullah:2018ykz, Gonzalez-Garcia:2018dep, Giunti:2019xpr, AristizabalSierra:2019ufd, Bischer:2019ttk, Canas:2019fjw, Babu:2019mfe, Denton:2020hop, Flores:2020lji}, 
axion-like particles~\cite{Dent:2019ueq}, 
light mediators and neutrino electromagnetic properties~\cite{Pospelov:2011ha, Harnik:2012ni, Pospelov:2012gm, Kosmas:2017tsq, Farzan:2018gtr, Boehm:2018sux, Denton:2018xmq, Cadeddu:2018dux, Billard:2018jnl, Dutta:2019eml, Cadeddu:2020nbr}, 
solar physics~\cite{Billard:2014yka, Cerdeno:2016sfi, Amaral:2020tga}, 
light sterile neutrinos~\cite{Formaggio:2011jt, Anderson:2012pn, Dutta:2015nlo, Canas:2017umu, Kosmas:2017zbh, Blanco:2019vyp, Miranda:2020syh}
and dark matter/dark sector models~\cite{Cui:2017ytb, Ge:2017mcq, Bertuzzo:2018itn, Dutta:2019nbn}.
Nevertheless, the potential of Skipper CCDs near the core of a commercial nuclear reactor, despite very promising, is still largely unknown.
Two main reasons for this stem from the scarcity of measurements on quenching factors at very low energies, and the characterization of backgrounds in these experiments. 
The latter is particularly important as Skipper CCD readout is relatively slow, making it very hard to implement active vetos to enhance signal to noise ratio.

A detailed study in this context seems very timely, particularly due to the proposal of the recently formed \emph{neutrino Interaction Observation with a Low Energy Threshold Array} (vIOLETA) collaboration: deploying a 1 to 10~kg Skipper CCD detector near the Argentinian commercial nuclear reactor Atucha~II~\cite{violeta, neutrino-2020-poster-1, neutrino-2020-poster-2, neutrino-2020-poster-3}. 
The present manuscript proposes to perform a study of physics potential, characterizing the relative importance of quenching, background modeling, systematics and statistics.
This may serve as a guide to future experiments as it identifies sensitivity bottlenecks and layouts where key improvements may be needed.

To exemplify the physics potential, we choose to analyze the sensitivity to the weak mixing angle.
The weak mixing angle is one of the key quantities related to the electroweak symmetry breaking of the standard model, as it relates the mass of the $W$ and $Z$ bosons, and its running in a given renormalization scheme such as $\overline{\rm MS}$ is an important consistency test of the standard model.
This is a challenging measurement which can serve as a proxy to demonstrate sensitivity to low scale new physics scenarios~\cite{Davoudiasl:2014kua}.
We expect the determination of $\sin\theta_W$ via CEvNS to be one of the major cornerstones in any precision CEvNS experiments~\cite{Akimov:2018ghi}.
Besides, there is no neutrino measurement of the weak mixing angle at scales below the GeV that proves to be competitive against other observations.

Finally, in our analysis we propose to use current data from reactor neutrino experiments, specifically Daya Bay~\cite{Adey:2019ywk}, to constrain the reactor neutrino flux.
This is very relevant given the current unresolved reactor antineutrino anomaly situation: the predicted neutrino fluxes do not agree with data from several experiments.
Our method would be more independent of the theoretical calculations and, more importantly, their estimated theory uncertainty.

\section{Coherent Elastic Neutrino-Nucleus Scattering}\label{sec:CEvNS}
Coherent elastic neutrino-nucleus scattering is a standard model process in which a neutrino scatters coherently off of a nucleus~\cite{Freedman:1973yd}.
If the momentum transfer is low enough, the wavelength of the (virtual) mediator can be comparable to the size of the nucleus.
The interaction then probes the nucleus as a whole, not distinguishing among neutrons and protons.
Neutral current interactions can be parametrized by the following effective Lagrangian
\begin{align*}
	\mathscr{L}_{\rm eff}^Z=4\sqrt{2}G_F J^\mu_ZJ_{\mu\,Z},
\end{align*}
where the neutral currents $J^\mu_Z$ are
\begin{align*}
	J^\mu_Z=\sum_{f=\nu,u,d} \left[\overline{f_L}\gamma^\mu \tau^3 f_L-\sin^2\theta_W \overline{f}\gamma^\mu Q_f f\right].
\end{align*}
Here, $G_F$ is the Fermi constant, $\tau^3$  denotes the third Pauli matrix, $\sin^2\theta_W\equiv 1-M_W^2/M_Z^2$ is the weak mixing angle ($M_{W,Z}$ are the masses of the weak gauge bosons), $Q_f$ is the electric charge of the fermions $f$ and the sum runs over $f=\nu,u,d$.
In this case, the differential standard CEvNS cross section can be written as 
\begin{align}\label{eq:recoil_SM}
  \frac{d\sigma^{\nu}}{dE_R}  &\simeq [{\cal Q}_V^{\rm SM}]^2 {\cal F}^2(E_R) \frac{G_F^2 m_N}{4\pi}   \left(1-\frac{m_N E_R}{2 E_\nu^2} \right) \, ,
\end{align}
where $m_N$ and $E_R$ are the mass and recoil energy of the struck nucleus and $E_\nu$ is the incoming neutrino energy.
As mentioned in the introduction, the weak charge of the nucleus is defined as 
\begin{align}
 {\cal Q}_V^{\rm SM} = N - (1-4 \sin^2\theta_W) Z,
\end{align}
where $N$ and $Z$ are the number of neutrons and protons in the target nucleus and ${\cal F}(E_R)$ is a form factor that parametrizes the coherence of the interaction. 
It is important to notice that this cross section is proportional to the square of the number of constituents, such as expected in a coherent scattering, with the modification introduced by  the $1-4 \sin^2\theta_W$ factor, coming from the different effective weak coupling of protons and neutrons. 

The form factor describes the coherence of the interaction.
Here, we will focus on the simplistic Helm form factor~\cite{Engel:1991wq, Lewin:1995rx, Harnik:2012ni},
\begin{align}\label{eq:FormFactor}
{\cal F}(E_R)&=3e^{-k^2s^2/2}\left[\sin(kr)-kr\cos(kr)\right]/(kr)^3
\end{align} 
where $k\equiv\sqrt{2 m_N E_R}$, $s \simeq 1$~fm is the nuclear skin thickness, $r=\sqrt{R^2-5s^2}$, and $R\simeq 1.14 \, (Z+N)^{1/3}$ is the  effective nuclear radius. 
This form factor parametrizes the internal structure of the nucleus as seen by the incoming neutrino. 

To have a better understanding of all those quantities, it is useful to plug in numbers relevant to our study into these formulas.
We will study the weak mixing angle sensitivity of an experimental setup consisting of a Skipper CCD detector deployed nearby a commercial nuclear reactor.
For typical neutrino energies at the MeV scale, the maximum nuclear recoil achievable for a nucleus $N$ is 
\begin{equation}
  E_R^{\rm max} = \frac{2E_\nu^2}{m_N+2E_\nu}\simeq 76~{\rm eV}\left(\frac{E_\nu}{\rm MeV}\right)^2\left(\frac{28~m_p}{m_N}\right),
\end{equation}
where $m_p\simeq 0.938$~GeV is the proton mass.
Here we can see the difficulty in detecting CEvNS with reactor neutrinos: the detector needs to be sensitive to energy depositions of 100~eV or so, to be able to take advantage of the large neutrino flux.
Conversely, the minimum neutrino energy for a given nuclear recoil is
\begin{equation}\label{eq:enumin}
  E_\nu^{\rm min}=\frac{1}{2}\left(E_R+\sqrt{E_R^2+2E_Rm_N}\right)\simeq1.1~{\rm MeV}\left(\frac{E_R}{100~{\rm eV}}\frac{m_N}{28m_p}\right)^{1/2}
\end{equation}
Finally, the typical momentum transfer $q$ for the recoil energies of interest is 
\begin{equation}
  \sqrt{-q^2}=\sqrt{2E_R m_N}\simeq 2.4~{\rm keV}\left(\frac{E_R}{100~{\rm eV}}\frac{m_N}{28 m_p}\right)^{1/2}
\end{equation}

\begin{figure}[t]
\includegraphics[width=.45\textwidth]{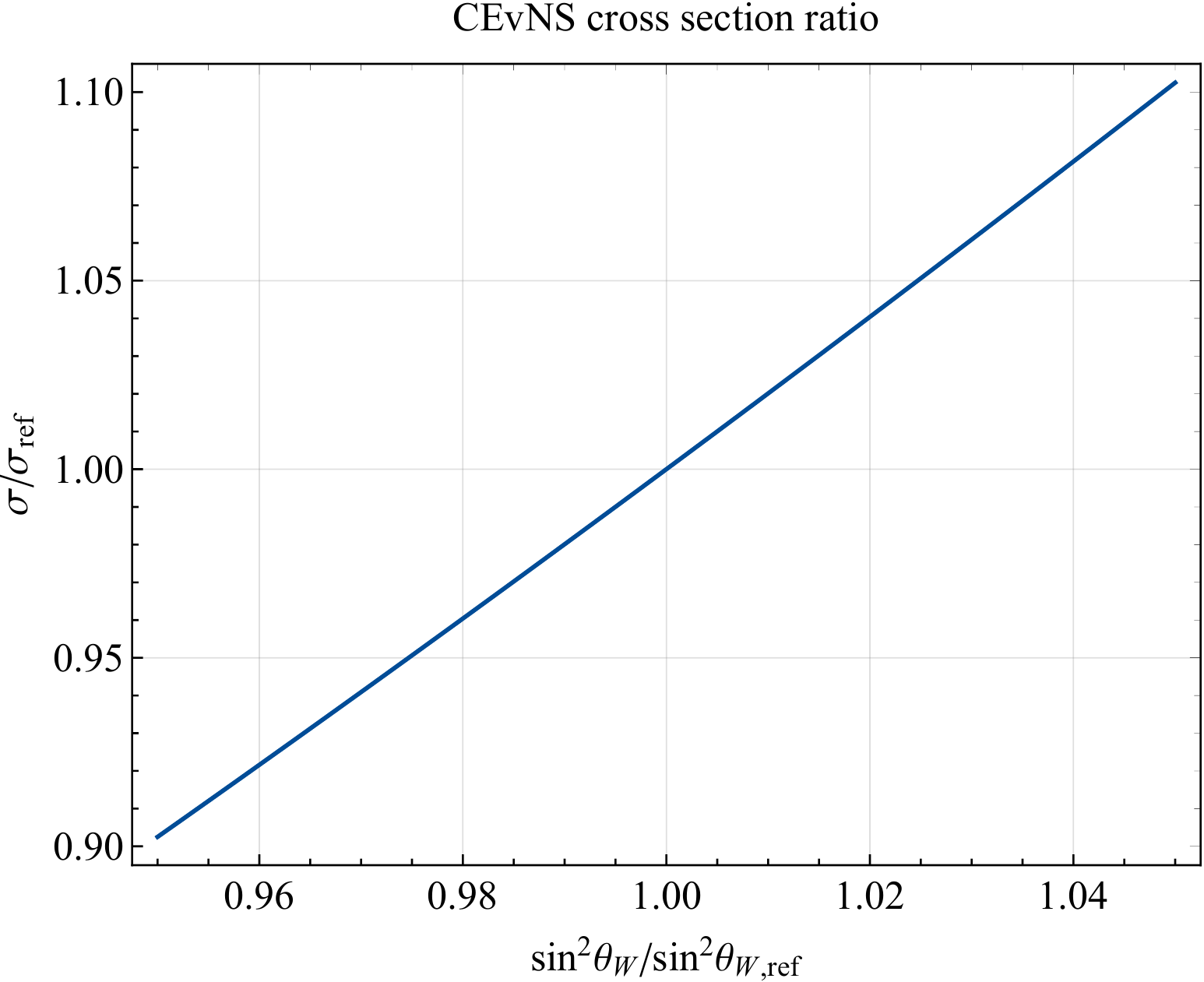}
\caption{Relative variation of the CEvNS cross section as a function of a change in the weak mixing angle $\sin^2\theta_W$. The reference value of the CEvNS cross section $\sigma_{\rm ref}$ is obtained with $\sin^2\theta_{W\!{\rm,ref}}=0.238$.
\label{fig:xsec-weak}}
\end{figure}

The form factor is more complicated to write in a useful form.
Nevertheless, noticing that 
\begin{equation}
  k s\simeq1.2\times 10^{-5} \left(\frac{E_R}{100~{\rm eV}}\frac{m_N}{28 m_p}\right)^{1/2}~~~{\rm and}~~~~
  kr\simeq3.2\times 10^{-5}\left(\frac{E_R}{100~{\rm eV}}\frac{m_N}{28 m_p}\right)^{1/2} ~~~\textrm{(for silicon)}
\end{equation} 
allows us to expand the form factor for small momentum transfer yielding ${\cal F}\sim 1 + \mathcal{O}(k^2s^2, k^2r^2)$.
Therefore, for this experimental setup, the form factor is very close to $\mathcal{F}(E_R)\sim1$ and we will approximate it to unity henceforth.
Using this approximation, we can approximate the differential cross section by 
\begin{equation}
  \frac{d\sigma}{dE_R}\simeq 2.3\times10^{-38} \left(1-\frac{2E_Rm_N}{E_\nu^2}\right) \left(\frac{{\cal Q}_V^{\rm SM}}{14}\right)^2\left(\frac{m_N}{28m_p}\right) \frac{{\rm cm}^2}{\rm eV}.
\end{equation}
Note however that $\mathcal{Q}_V^{\rm SM}$ depends on the weak mixing angle.
The last expression we will present here draws upon the fact that $\sin^2\theta_W\simeq 0.238$ is close to $1/4$. 
By writing $\mathcal{Q}_V^{\rm SM}$ around $\sin^2\theta_W=0.25(1-\epsilon)$ we obtain
\begin{equation}
  [\mathcal{Q}_V^{\rm SM}]^2= N^2\left(1-2\frac{Z}{N} \epsilon+\frac{Z^2}{N^2}\epsilon^2\right).
\end{equation}
For example, from the above formula we expect a 1\% change in $\sin^2\theta_W$ to correspond roughly to a 2\% change in the cross section, see Fig.~\ref{fig:xsec-weak}.

Finally, a few words regarding the weak mixing angle. 
The definition of the weak mixing angle and its renormalization group running depends on the renormalization scheme.
In this work we adopt the $\overline{\rm MS}$ renormalization scheme, in which the weak mixing angle for very low momentum transfer is predicted to be $\sin^2\theta_W (Q^2)=0.23867$ from the measurement of the same angle at the Z boson pole~\cite{Zyla:2020zbs}. 
In the standard model, $\sin^2\theta_W$ does not run considerably below $Q^2\equiv-q^2\sim(100~{\rm MeV})^2$ or so.
Therefore, we will assume this value in the following as input in order to estimate the sensitivity of our experimental setup to the weak mixing.

\section{Analysis}
\label{sec:analysis}

The experimental setup we consider here consists of a Skipper CCD detector, deployed at 12~meters from the main core of a nuclear reactor like the Atucha~II reactor complex, located at the province of Buenos Aires, Argentina.
Throughout our analysis, the detector is assumed to have a fiducial mass of 1~kg.
Later, we will comment on the physics case of a 10~kg detector.
The core is assumed to have a thermal power of 2~GW, emitting about $\sim10^{28}$ electron antineutrinos per year. 
We will assume a data taking period of 3 years.

Regarding backgrounds, it is expected that the Compton interactions from high energy photons and nuclear interactions from high energy neutrons will be the dominant  contributions for the recoil energies of interest. Both processes are well predicted by state-of-the-art particle simulators, but should be analyzed for a specific passive shield configuration to have their impact properly quantified.
Therefore, this experimental setup depends crucially on background measurements during reactor-off periods, which are made as short as possible in commercial nuclear reactors.
To be realistic, we assume an average reactor-off time of 45 days per year.
For the background estimation, we call attention to the measured background in the CONNIE experiment of about 10~kdru (1 \emph{differential rate unit}, or dru, is 1 event per day-keV-kg)~\cite{Aguilar-Arevalo:2019jlr}. 
New studies on the CCD used by CONNIE show that some  of its background can be produced by a partial charge collection layer in the back of the sensor~\cite{pcc-2019}. 
Large charge packets created in this highly doped region are observed as lower energy events since some of the carriers are lost by recombination. 
This problem can be eliminated by a back side processing of the sensor as explained and tested in the same publication. 
The CCD detector of CONNIE was located right outside the reactor dome with almost no virtual overburden besides the passive shield. 
In the scenario considered here, the detector is inside the dome providing some extra reduction to cosmic background. 
Therefore, we will assume a baseline background of 1~kdru, with flat deposited energy spectrum.
We will study the impact of higher or lower background rates, as well as the impact of the background spectral shape~\cite{Bowen:2020unj}.
Moreover, we will consider the reactor-on background to be negligible, although this needs to be determined by detailed simulations and mandatorily avoided by adding shielding between the detector and the reactor core.

Skipper CCD sensors have shown very low leakage current contributions~\cite{Abramoff:2019dfb, Barak:2020fql} that allow to explore ionizations up to one or two electrons. 
Nevertheless we assume a conservative minimum  of ionization energy of 15 eV, corresponding on average to four ionized electrons.
It should be noted that the translation from nuclear recoil energy to ionization energy, which is what is detected by Skipper CCDs, is encoded by the \emph{quenching factor}.
At such low recoils, the quenching factor is not well-known, and measurements to determine it are planned using the Skipper sensors following the procedure presented in Ref.~\cite{Chavarria:2016xsi}.
This is critical in reactor neutrino CEvNS, as the neutrino flux grows considerably at lower neutrino energies.
We will estimate the effect of two different quenching functions and its uncertainties on the experimental sensitivity.
Our benchmark will assume the parametrization from Ref.~\cite{Aguilar-Arevalo:2019zme} of the measurement performed in Ref.~\cite{Chavarria:2016xsi}.
As the expected neutrino ionization energy spectrum is fairly broad, and the reconstruction of 
the ionization energy resolution in Skipper CCDs is excellent 
(essentially only affected by the silicon absorption, which can be estimated through the fano factor~\cite{Rodrigues2020}),  we do not consider any ionization energy smearing here.
Note however that this does not help in reconstructing the incoming neutrino energy.
We bin the simulated data in 50~eV ionization energy bins.

Finally, several systematic uncertainties may affect the physics reach of these experiments.
Among the most important ones is the uncertainty on the reactor neutrino flux.
Several theoretical estimates of this flux have been performed~\cite{Mention:2011rk, Huber:2011wv}.
Although the calculation of these fluxes involve thousands of beta branches and forbidden decays, they achieve a remarkable systematic error of about 2\%.
Nevertheless, these calculations do not agree with the measured reactor antineutrino flux via the inverse beta decay process, giving rise to a $3\sigma$ anomaly, dubbed \emph{the reactor antineutrino anomaly}.
The reason for this discrepancy is still unknown, ranging from too aggressive uncertainties~\cite{Hayes:2013wra, Hayes:2016qnu, Sonzogni:2017wxy} to new physics beyond the standard model~\cite{Dentler:2018sju, Diaz:2019fwt, Boser:2019rta}.
On top of that, theoretical calculations fail to predict a relatively large feature in the reactor neutrino spectrum around neutrino energies of 5 MeV (see e.g. Ref.~\cite{Bak:2018ydk}). 
In view of that, we will estimate the experimental sensitivity to $\sin^2\theta_W$ using the flux determination from the Daya Bay experiment~\cite{Adey:2019ywk} instead of relying on theoretical calculations of the neutrino flux and the corresponding systematic uncertainties.
Although the former has larger uncertainties, it is a data-driven approach, and thus it is more robust than theoretical estimates.
Later, we will show that the flux uncertainty does not play a role as important as other factors such as background rate and quenching factor determination.

The expected signal event rate, for each fuel isotope, measured by the Skipper CCD detector can be obtained by convolving the neutrino flux with the CEvNS cross section, taking into account the quenching factor, namely,
\begin{align}\label{eq:signal}
	n_{aj}^q = \frac{W}{\sum_{q'} (f_{q'} e_{q'})}\int_\textrm{bin $a$} \!\!\!\!\!\!\!\!dE_I  \frac{1}{Q(E_I)}\left(1-\frac{E_I}{Q(E_I)}\frac{dQ(E_I)}{dE_I}\right) \int_\textrm{bin $j$} \!\!\!\!\!\!\!\!dE_\nu\, \frac{d\phi_{\bar{\nu}_e}^q}{d E_\nu}\left.\frac{d\sigma_{\rm CE\nu NS}}{dE_R}\right|_{E_R=E_I/Q(E_I)}.
\end{align}
We denote the ionization energy by $E_I$, and the nuclear recoil energy by $E_R$.
$W$ is the total reactor power.
For convenience, we have assigned three indices to $n^q_{aj}$.
The upper index $q$ indicates the fuel isotope, that is, $^{235}$U, $^{239}$Pu, $^{238}$U and $^{241}$Pu.
Associated to that is the 
relative rate per fission $f_{q'}$ and energy released per fission $e_{q'}$ for isotope $q'$.
The first lower index corresponds to the ionization energy bin, while the second one refers to the incoming neutrino energy integration interval.
Including the latter is important as it allows us to properly take care of flux uncertainties.
The quenching function is $Q(E_I)=E_I/E_R$.
Note that the terms in the outer integral come from the Jacobian when changing the integration variable from $E_R$ to $E_I$.
The intervals of integration should be appropriate for the binning and also respect Eq.~\eqref{eq:enumin}.
The neutrino flux per fission for each isotope is given by $\phi_{\bar\nu_e}^q$.
For energies above the inverse beta decay threshold, $E_{\nu}>1.8$~MeV, we use the antineutrino fluxes measured by Daya Bay~\cite{Adey:2019ywk}. In their analysis, Daya Bay uses the theoretical estimates from Refs.~\cite{Mention:2011rk, Huber:2011wv} as inputs for $^{238}$U and $^{241}$Pu and extracts the fluxes for the two
isotopes with the largest relative rates per fission, $^{235}$U and $^{239}$Pu, from their measurements. For energies lower than the inverse beta decay threshold we will use the flux estimate of Vogel and Engel~\cite{Vogel:1989iv}. We will assume an uncertainty of 5\% in the relative rate per fission of the flux for every isotope.

The sensitivity over $\sin^{2}\theta_{W}$ is determined 
by minimizing the following
$\chi^2$ 
for the reactor-on running,
\begin{align}\label{eq:chi2}
    \chi^2=(\mathbf{d}-\mathbf{t})^T\mathbf{C}^{-1}(\mathbf{d}-\mathbf{t})+\sum_a^\textrm{bins}\left(\frac{\alpha^B_a}{\sigma^B_a}\right)^2+\left(\frac{\alpha_W}{\sigma_W}\right)^2+\sum_q^\textrm{isotopes}\left(\frac{\alpha_q}{\sigma_q}\right)^2+\left(\frac{\alpha_Q}{\sigma_Q}\right)^2.
\end{align}
There are several quantities in this expression
that need to be defined. 
Let us start with the last four terms which correspond to the Gaussian distributions (also sometimes referred as Gaussian priors)
that parametrize the systematic uncertainties of the present experiment.
The first systematic is related to the background determination.
During reactor-off running, the experiment will determine the background in each bin. 
The relative statistical uncertainty of this determination is what enters in $\sigma^B_a$.
The second systematic is on the total reactor power $W$. 
The third systematic corresponds to the relative rate per fission $f_q$, which depends on the isotope composition.
The last one is on the quenching factor $Q(E_I)$.
The $\alpha$'s are pull parameters that will be minimized for each value of $\sin^2\theta_W$ according to the profiling method.

The approach adopted here, regarding the determination of the background, is most conservative.
In principle, one can also adopt a more aggressive approach.
 For example, if the background can be well modeled by MonteCarlo simulations, say by a polynomial function, the reactor-off running would determine the uncertainties on the coefficients of such function, which would then be used in Eq.~\eqref{eq:chi2} (the implementation of the pull parameters would change accordingly).

Now we move on to the first term of the $\chi^2$. 
Here, $\mathbf{d}$ and $\mathbf{t}$ are vectors of data (Asimov data with $\sin^2\theta_W=0.238$) and theory (where $\sin^2\theta_W$ is allowed to vary) events, respectively. 
These vectors can be constructed by summing up all neutrino energies and isotope contributions in the signal rate in Eq.~\eqref{eq:signal}, as well as background rates.
\begin{equation}
  {\rm t_a,d_a} = \sum_{j,q} f_q n_{aj}^q + B_a,
\end{equation}
where $B_a$ is the background in ionization energy bin $a$.
Note that the theory prediction, $\mathbf{t}$, should also properly include the pull parameters $\alpha$.
To do so, the following substitutions should be made 
\begin{align}
  f_q\to(1+\alpha_q)f_q,\qquad B_a\to(1+\alpha^B_a)B_a,\qquad W\to(1+\alpha_W)W, \qquad Q(E_I)\to(1+\alpha_Q)Q(E_I).
\end{align}
Note that the relative rate per fission should sum up to 1, and therefore $\sum_q (1+\alpha_q)f_q = 1$.
For the reference values we use $f_{235}:f_{238}:f_{239}:f_{241} = 0.55 : 0.07 : 0.32 : 0.06 $.

The covariance matrix $\mathbf{C}$ parametrizes  correlated and uncorrelated systematic uncertainties in the determination of the flux, as well as statistical uncertainties.
We adopt 
\begin{align}
    \mathbf{C}_{ab} = \sum_{ij}^{\rm energies}\,\sum_{q}^{\rm isotopes} n_{aj}^q n_{bk}^q V_{jk}^q+ {\rm d}_a\delta_{ab},
\end{align}
where systematic and statistical uncertainties are encoded in the first and second terms, respectively. 
Note that the Daya Bay collaboration only provides this correlation matrix for the 
isotopes with the largest rate per fission, namely, $^{235}$U and $^{239}$Pu, and for neutrino energies above 1.8~MeV (which is the threshold for inverse beta decay). 
In this case, $V_{ij}^q$ are taken from Daya Bay's flux covariance matrix~\cite{Adey:2019ywk} and correlate the uncertainties in the measured spectrum.
For the flux below 1.8~MeV, we use a single neutrino energy bin, and we assume an uncorrelated spectral uncertainty of 5\%.
Moreover, for the subdominant rate per fission of the isotopes $^{238}$U and $^{241}$Pu, we use a 5\% bin-to-bin uncorrelated systematic uncertainty on the flux.
Given that the relative rate per fission of these isotopes are assumed to be 0.07 and 0.06, such systematic uncertainties play a small role in the analysis.

The numerical values for the priors used can be found in Table.~\ref{tab:systematics}.
Note that the background uncertainty is derived from the reactor-off period.
We assume the background rate to be 1~kdru, flat in ionization energy, an exposure of 45 days per year, for 3 years, and a detector mass of 1~kg.
This yields a statistical determination of the background rate of 1.2\% in each 50~eV ionization energy bin. 

\begin{table}[t]
\begin{center}
\begin{tabular}{ |l | c | c | c | c |}\hline\hline
Systematic~   & ~Background~ & ~Reactor power~& ~Relative rate per fission~ & ~Quenching~\\ \hline
Symbol   &   $\sigma_{a}^B$   &   $\sigma_{W}$   &   $\sigma_{q}$   &   $\sigma_{Q}$   \\
Value   &   $1.2\%$   &   $0.5\%$   &   $5\%$   &   $0-25\%$   \\ \hline \hline
\end{tabular}
\end{center}
\caption{Systematic uncertainties used in our analysis. See text for details.\label{tab:systematics}}
\end{table}

\section{Results}
\label{sec:res}

\begin{figure}[h]
\includegraphics[width=.45\textwidth]{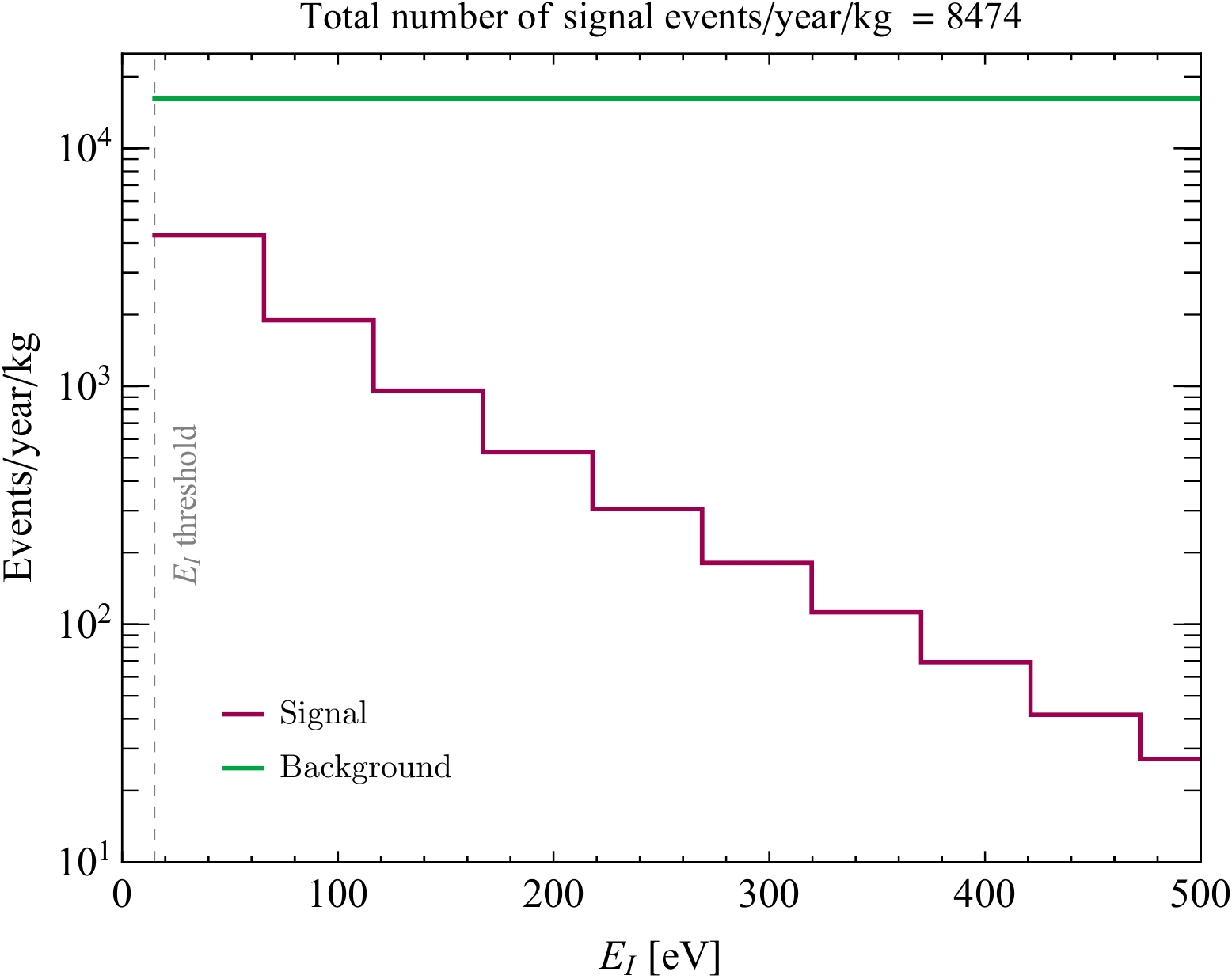}
\includegraphics[width=.465\textwidth]{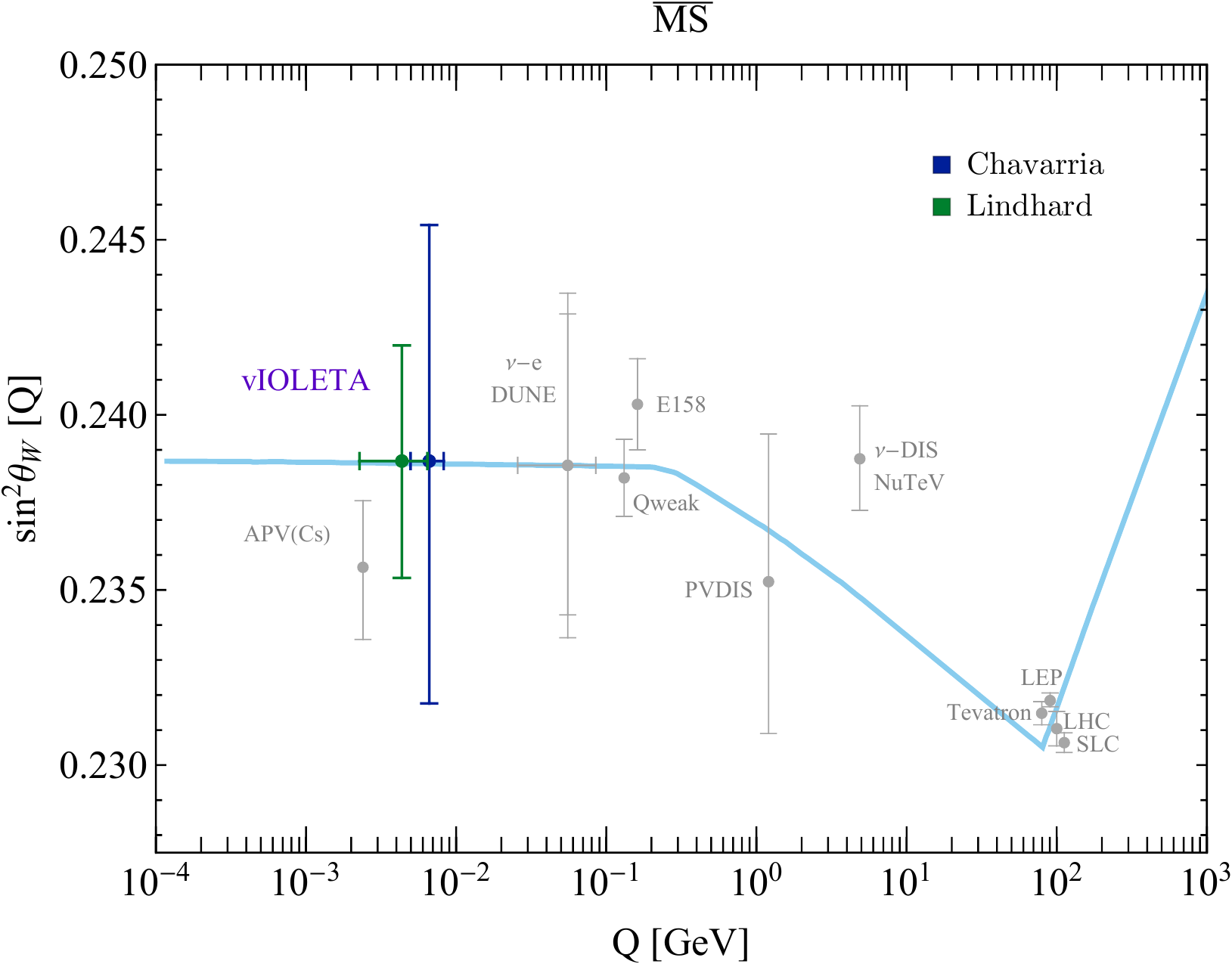}
\caption{Left: Spectral rate of signal (red histogram) and background (green) events for our benchmark setup (see Table~\ref{tab:benchmark}). The vertical dashed line shows the ionization energy threshold of Skipper CCDs. Right:  Sensitivity of our experimental setup to $\sin^2\theta_{W}$ compared with different experiments~\cite{Zyla:2020zbs, deGouvea:2019wav} in the $\overline{\rm MS}$ renormalization scheme. The two colored data points correspond to two different quenching factors, Lindhard (green) and Chavarria (dark blue). No uncertainty on the quenching was considered. 
\label{fig:Event}}
\end{figure}

We now proceed to the results of our simulation.
Our benchmark setup is summarized in Table~\ref{tab:benchmark}.
We  assume the quenching factor to be the one measured in Ref.~\cite{Chavarria:2016xsi} and parametrized in Ref.~\cite{Aguilar-Arevalo:2019zme}.
The quenching models and the impact of systematic uncertainties will be discussed in detail below.
The signal and background event rates spectra for this benchmark are presented in the left panel of Fig.~\ref{fig:Event}.
We will perform studies on variations of this scenario to understand the role of different aspects on the experimental sensitivity.
The sensitivity to the weak mixing angle in our benchmark scenario is shown in the right panel of Fig.~\ref{fig:Event} for two different assumptions on the quenching factor which will be discussed below, the Lindhard and the Chavarria models, without systematic uncertainties on the quenching. 
Together with our estimate, we also present current measurements (taking into account the running of $\sin^2\theta_W$)~\cite{Zyla:2020zbs} and a forecast for the measurement on the DUNE near detector complex~\cite{deGouvea:2019wav}. 

\begin{table}[t]
\begin{center}
\begin{tabular}{ |l | c |}\hline\hline
Detector mass  & 1~kg\\ \hline
Distance to reactor core  & 12 meters\\ \hline
Core thermal power  & 2~GW$_{\rm th}$\\ \hline
Exposure  & 3 years\\ \hline
Reactor off time  & 45 days per year\\ \hline
Background  &1~kdru, flat in $E_I$ \\ \hline
Quenching factor & Chavarria~\cite{Chavarria:2016xsi} \\ \hline\hline
\end{tabular}
\end{center}
\caption{Benchmark experimental setup considered here.\label{tab:benchmark}}
\end{table}

The precision achieved by this benchmark setup is $1.4\%$ for Lindhard and $2.8\%$ for Chavarria quenching, at mean momentum transfers $\langle Q^2\rangle$ of 4.3~MeV and 6.6~MeV, respectively. 
The mean momentum transfer is different because the minimum detectable recoil energy depends on the quenching.
As we can see, under reasonable assumptions, this experimental setup would determine the weak mixing angle with a precision similar or slightly worse than atomic parity violation (APV), and comparable precision to several other determinations, including the future DUNE experiment~\cite{deGouvea:2019wav}.
The competitiveness of such a small experiment is quite remarkable.
The measurement proposed here would be one of the very few determinations of the weak mixing angle with neutrinos, which is particularly important given the discrepancy between measurement and theory prediction observed in the NuTeV experiment~\cite{Zeller:2001hh}.
As a side comment, by using the LEP determination of the weak mixing angle and the theoretical prediction for the reactor antineutrino flux, this setup could  probe the reactor anomaly at the $3-6\%$ precision.
To estimate how robust this measurement is and how it depends on our assumptions, we proceed to a study of how each ingredient affects our analysis.
We will study the role of quenching factor parametrization and uncertainties, as well as background rate and shape.

\subsection{Quenching factor}

As mentioned before, the quenching factor characterizes the fraction of the recoil energy transformed into observable ionization energy. 
Measurements of the quenching factor in the case of silicon have been previously performed~\cite{Chavarria:2016xsi}. 
Low energy measurements disagree with the theoretical prediction obtained using the Lindhard model~\cite{lindhard} for the quenching.
Given this discrepancy, we will study two cases for the quenching factor.

\textbf{Lindhard model.}
We start by describing the implementation of the widely used Lindhard model for the quenching factor at low energies.
We perform a polynomial fit to the Lindhard  quenching model given in Ref.~\cite{Mei:2007jn}, obtaining the following functional form
\begin{equation}
    Q(E_{I}) = a_{0} + a_{1} E_{I} + a_{2} E_{I}^{2} + a_{3} E_{I}^{3} + a_{4} E_{I}^{4} + a_{5} E_{I}^{5} + a_{6} E_{I}^{-1} + a_{7} E_{I}^{-2},
\end{equation}
where $a_0=0.173$, $a_{1}=0.307~\keV^{-1}$, $a_{2}=-0.856~\keV^{-2}$, $a_{3}=1.684~\keV^{-3}$, $a_{4}=-1.801~\keV^{-4}$, $a_{5}=0.783~\keV^{-5}$, $a_{6} = 5.9\times 10^{-4}~\keV$ and $a_{7}=3.3\times 10^{-6}~\keV^{2}$. 
Note that, within this parametrization, the Skipper CCD threshold of $E_I^{\rm min} = 15$~eV would translate to a recoil of $E_R^{\rm min} = E_I^{\rm min} /Q = 65$~eV.
From Eq.~\eqref{eq:enumin}, the minimum neutrino energy to produce such recoil is about 0.9~MeV. 
Therefore, assuming the Lindhard quenching model, a large portion of the reactor antineutrino flux, below the inverse beta decay threshold, can lead to observable recoils in our experimental setup.

\textbf{Chavarria model.}
The measurements performed to determine the quenching factor at low energies~\cite{Chavarria:2016xsi} can be parametrized by the following ratio of polynomials
\begin{align}
	Q(E_I)=\frac{p_3 E_I+p_4 E_I^2+E_I^3}{p_0 +p_1 E_I+p_2 E_I^2}.
\end{align}
A fit to the data determines $p_0=56\  \keV^3$, $p_1 = 1096\ \keV^2$, $p_2 = 382\ \keV$, $ p_3 = 168\ \keV^2$ and $p_4 = 155\ \keV$. 
This parametrization, which we will refer to as the \emph{Chavarria} model, has been used by the CONNIE experiment in beyond standard model physics searches~\cite{Aguilar-Arevalo:2019zme}. 
Given that the quenching factor has only been measured down to ionization energies of about $60$~eV and the measurement is not in agreement with the theoretical predictions of Lindhard, it is not completely clear what should be assumed for the quenching below that energy.
Thus, we use this parametrization as an alternative to the Lindhard quenching factor described above.

One important feature of the Chavarria quenching is that, for small enough $E_I$, the recoil energy $E_R = E_I/Q$ becomes constant.
Therefore, it becomes very hard to detect low recoils, as the ionization energy shrinks very quickly for small $E_R$.
Plugging in numbers, the 15~eV ionization energy threshold translates into a recoil energy of 425~eV, which in turn gives us a minimum neutrino energy of $2.3$~MeV, see Eq.~\eqref{eq:enumin}.
Thus, compared to the Lindhard quenching, the Chavarria quenching will significantly decrease statistics. 
Besides, a systematic uncertainty on the Chavarria quenching will change the minimum neutrino energy considerably and therefore significantly affect the rate of signal events.
We emphasize however that this assumption about the quenching is very conservative and future measurements will clarify the quenching below 60~eV of ionization energy.

\begin{figure}[t]
\includegraphics[width=.4\textwidth]{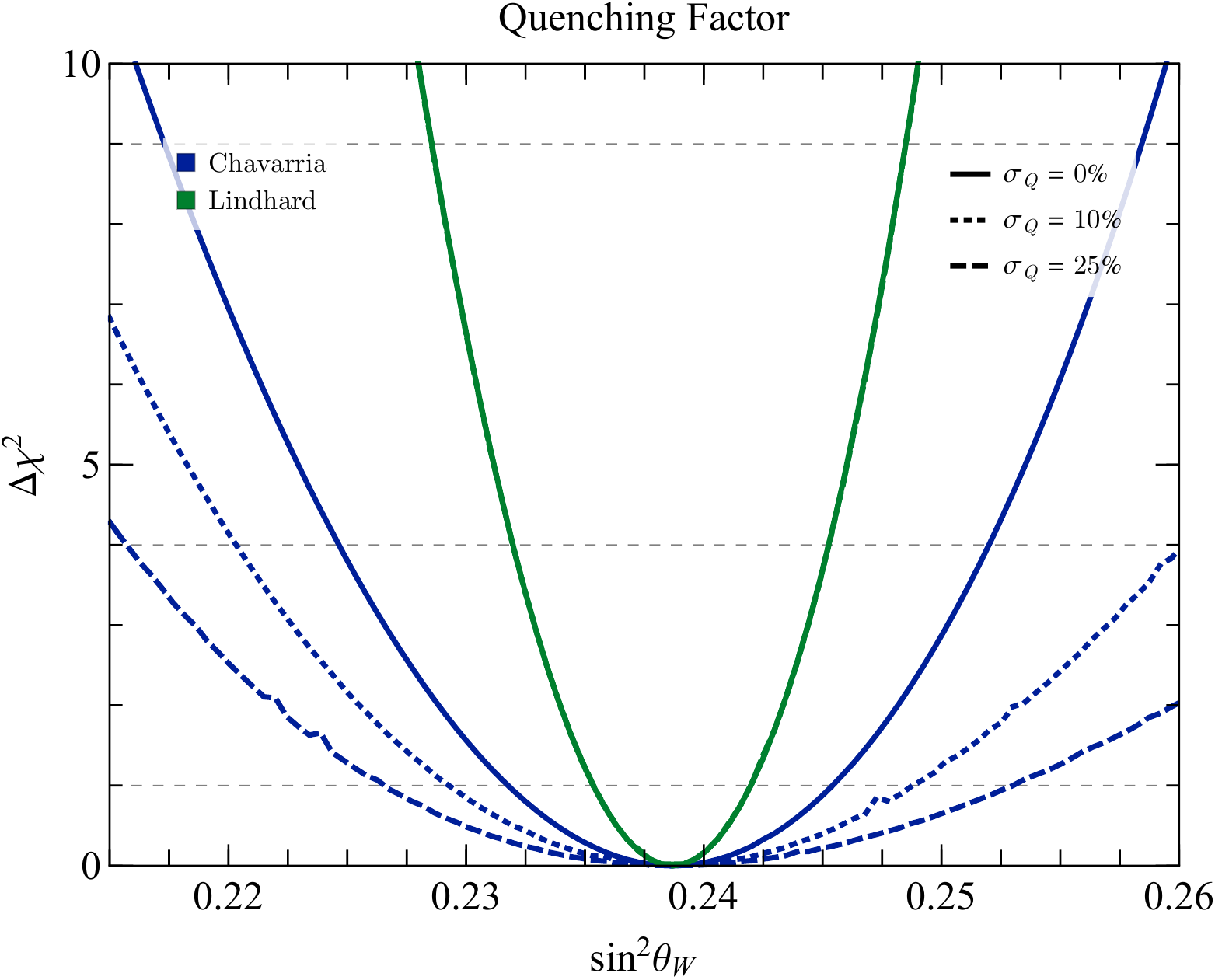}
\caption{Benchmark sensitivity to $\sin^2\theta_W$ for  Lindhard (green) and Chavarria (dark blue) quenching factor models, for systematic uncertainties of 0\% (solid), 10\% (dotted) and 25\% (dashed). We indicate 1$\sigma$, 2$\sigma$, and 3$\sigma$ precision ($\Delta\chi^2=1,4,9$) by the gray dashed horizontal lines.
\label{fig:ComparisonQuenching}
}
\end{figure}

\textbf{Quenching factor systematics.}
To understand the impact of systematic uncertainties we  evaluate the $\sin^2\theta_W$ sensitivity under several assumptions and present the results in Fig.~\ref{fig:ComparisonQuenching}.
The assumptions regarding quenching and its systematic uncertainties are the following: Lindhard (green) or Chavarria (dark blue) models, with 0\% (solid), 10\% (dotted) and 25\% (dashed) overall normalization uncertainties in the quenching (note that for Lindhard these lines are all on top of each other).
As discussed, the Lindhard quenching yields a considerably superior result compared to the Chavarria model due to statistics.
Moreover, in the Lindhard model a variation in the quenching does not change the rate of signal events by much, due to the very low thresholds of Skipper CCD detectors.
For any of the quenching uncertainties, the Lindhard quenching would allow for a $1.4\%$ precision measurement of the weak mixing angle.
For the more conservative quenching model of Chavarria, we find that the precision on the mixing angle is significantly decreased to be $2.8\%$, $4.2\%$ and $6.0\%$ for a 0\%, 10\% and 25\% systematic uncertainty, respectively.
This dramatic effect is simply due to the fact that variations in the quenching lead to significant variations in the signal rates, as much more or much less neutrino flux becomes detectable.
We conclude, therefore, that the determination of the quenching factor at low ionization energies is crucial for the physics case of an experiment measuring  CEvNS from reactor neutrinos with a Skipper CCD detector.

\subsection{Backgrounds}

In CEvNS, unless the recoiling nucleus is tracked and reconstructed, which is extremely challenging, the typical experimental signature is quite simple: just a small deposition of energy in the detector.
Because of that, background rejection is a difficult task in CEvNS detectors. 
This is particularly true for Skipper CCDs.
Due to their slow readout, not even active vetos may be used to reject the cosmic induced background.
Background rates and measurements are therefore crucial to the success of the experimental setup considered here.
We will quantify this statement in the following by analyzing the dependence of the experimental sensitivity on the background rate, modeling, and shape.
Our benchmark assumes reactor-off running time of 45 days per year which serves to determine the background rate.
In the background determination, we simply use the uncorrelated bin-to-bin statistical uncertainty from the reactor-off data taking period.
A background model could significantly improve the experimental sensitivity, but that would require a dedicated background study, which is beyond the scope of this paper.
For concreteness, we will use the Chavarria quenching as it will result in a more conservative sensitivity estimate, but without quenching systematic uncertainty.

\begin{figure}[h]
\includegraphics[width=.49\textwidth]{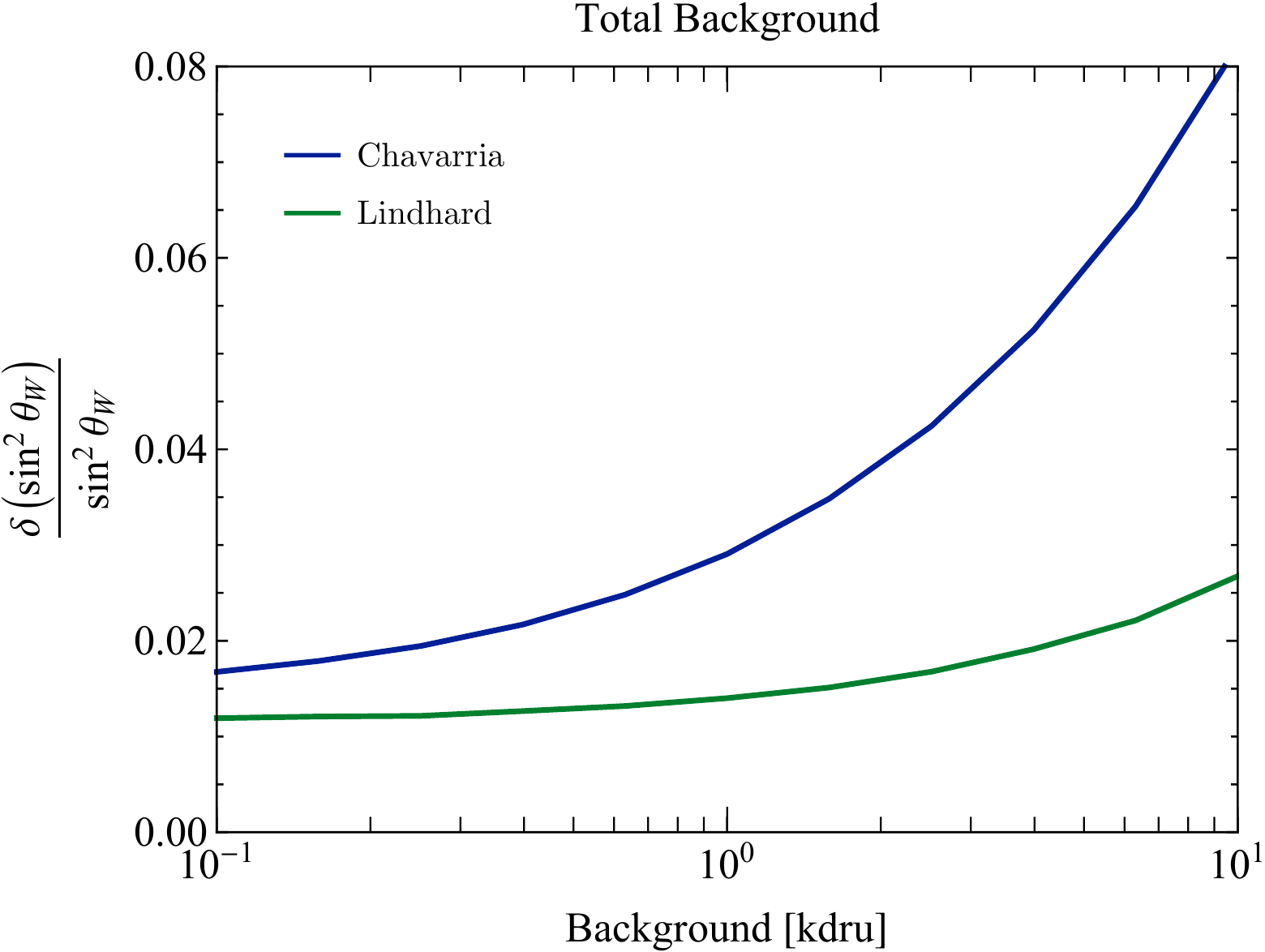}
\includegraphics[width=.45\textwidth]{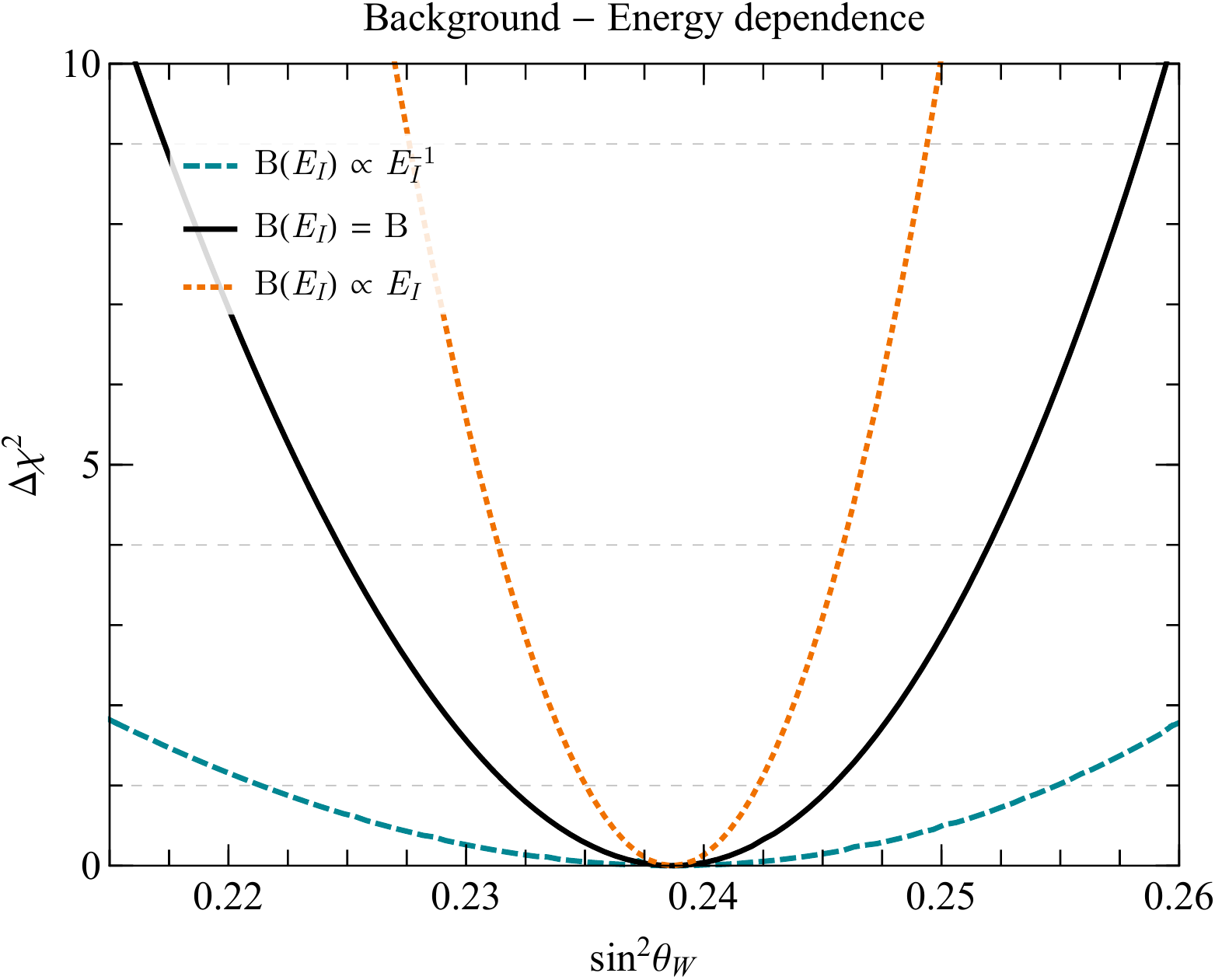}
\caption{Left: Precision on the determination of $\sin^2\theta_W$  as a function of background rates, assuming a flat background spectrum, for Chavarria (blue) and Lindhard (green) quenching scenarios.
Right: Benchmark sensitivity to $\sin^2\theta_W$ for different background spectral shapes:  flat (black solid), $E_I^{-1}$ (dark cyan dashed) and $E_I$ (dotted orange).}
\label{fig:back_norm}
\end{figure}

First, we investigate the effect of increasing or decreasing the background rate. 
In the left panel of Fig.\ref{fig:back_norm}, we present relative $1\sigma$ determination of $\sin^2\theta_W$ as a function of the total background rate for both Chavarria (blue line) and Lindhard (green line) quenching factors.  
We observe that, in the case of Chavarria quenching, the measurement of the weak mixing angle is highly sensitive on the total background. 
For instance, if the background is 10~kdru, that is, ten times larger than in our benchmark scenario, the precision on $\sin^2\theta_W$ gets worse from $2.8\%$ to $8\%$, while for  $0.1\ {\rm kdru}$ background rate the precision is enhanced to $1.6\%$. 
The  dependence on the total background rate is less pronounced for Lindhard quenching, as the signal rate in this case is much higher, and therefore the experimental sensitivity is less affected by backgrounds around 1~kdru.

To see how the shape of the background changes the sensitivity, in the right panel of Fig.~\ref{fig:back_norm}, we perform the analyses for three representative spectral shapes in ionization energy: flat (black), proportional to $E_I^{-1}$ (dashed dark cyan), and linear in $E_I$ (dotted orange). 
In doing these studies, we have kept the background rate fixed at 1~kdru in the energy range $E_I\in [15,675]$~eV.
As already pointed out in Ref.~\cite{Bowen:2020unj}, if the background grows at low energies, the experimental sensitivity is significantly decreased.
This is because in that case, the background has a shape very similar to the signal, see Fig.~\ref{fig:Event}.
Conversely, if the background shrinks at low ionization energy, the experimental sensitivity would be increased.
From this, we can appreciate the relevance of properly modeling and mitigating the background rate in experiments employing Skipper CCD detectors.

\subsection{Detector mass}
\begin{figure}[h]
\includegraphics[width=.45\textwidth]{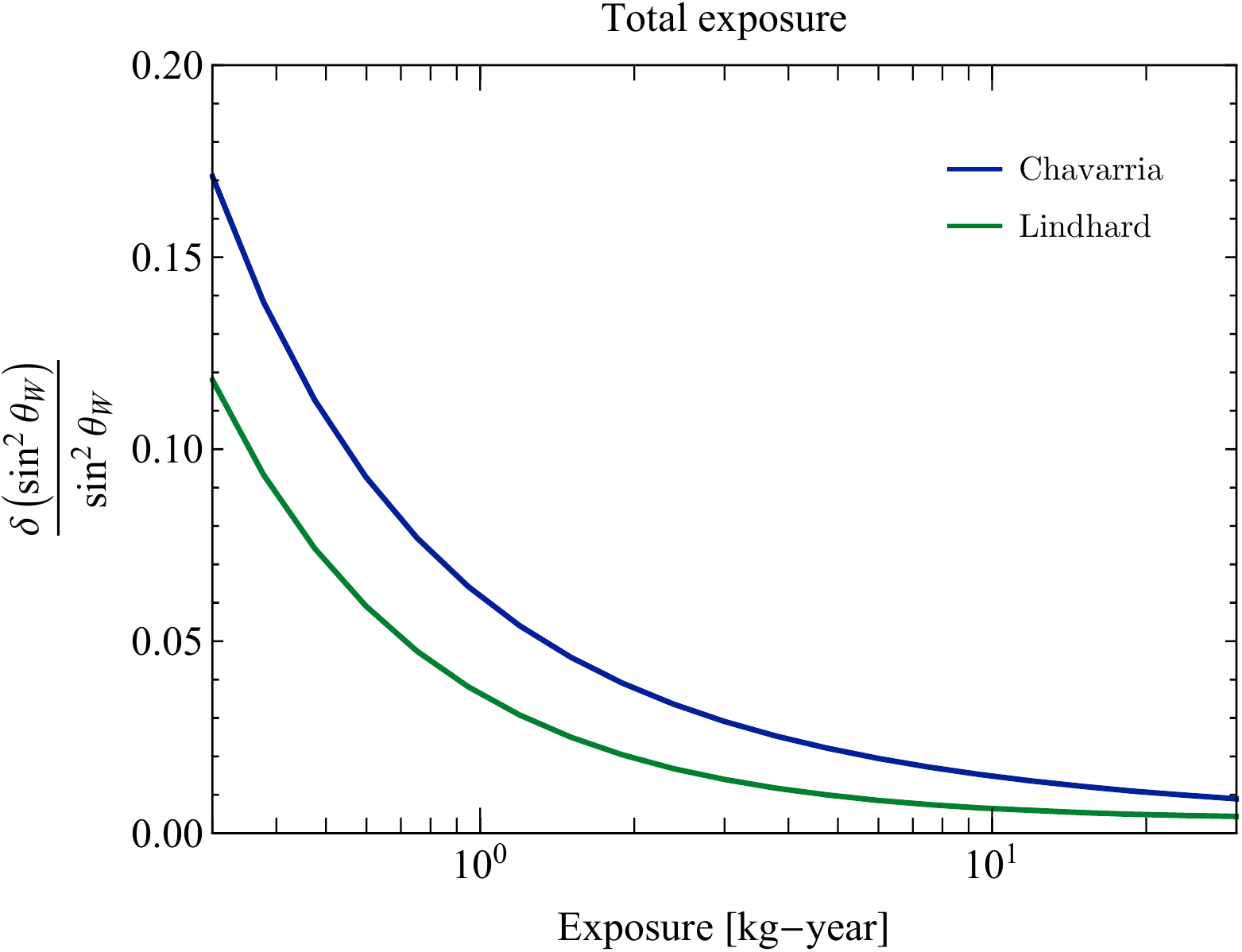}
\caption{Precision on the determination of $\sin^2\theta_W$ as a function of the total exposure for Chavarria (blue) and Lindhard (green) quenching scenarios.}
\label{fig:exposure}
\end{figure}

We also perform a study the dependence of the $\sin^2\theta_W$ determination as a function of the total exposure.
This helps to understand if the sensitivity is systematic or statistically limited.
In Fig.~\ref{fig:exposure}, we show, for our benchmark scenario, how the $1\sigma$ uncertainty on $\sin^2\theta_W$ depends on the total exposure for Chavarria (blue line) and Lindhard (green line) quenching.
In the first case, going from 3 kg-year to 30~kg-years of exposure, the precision on the weak mixing determination could go from 2.8\% to 0.9\% evidencing that the experiment sensitivity is not systematics limited, and thus a larger detector or longer exposure time would yield significantly better physics measurements.
We observe a similar pattern for the Lindhard quenching case.

\subsection{Reactor antineutrino flux uncertainty}
\begin{figure}[h]
\includegraphics[width=.45\textwidth]{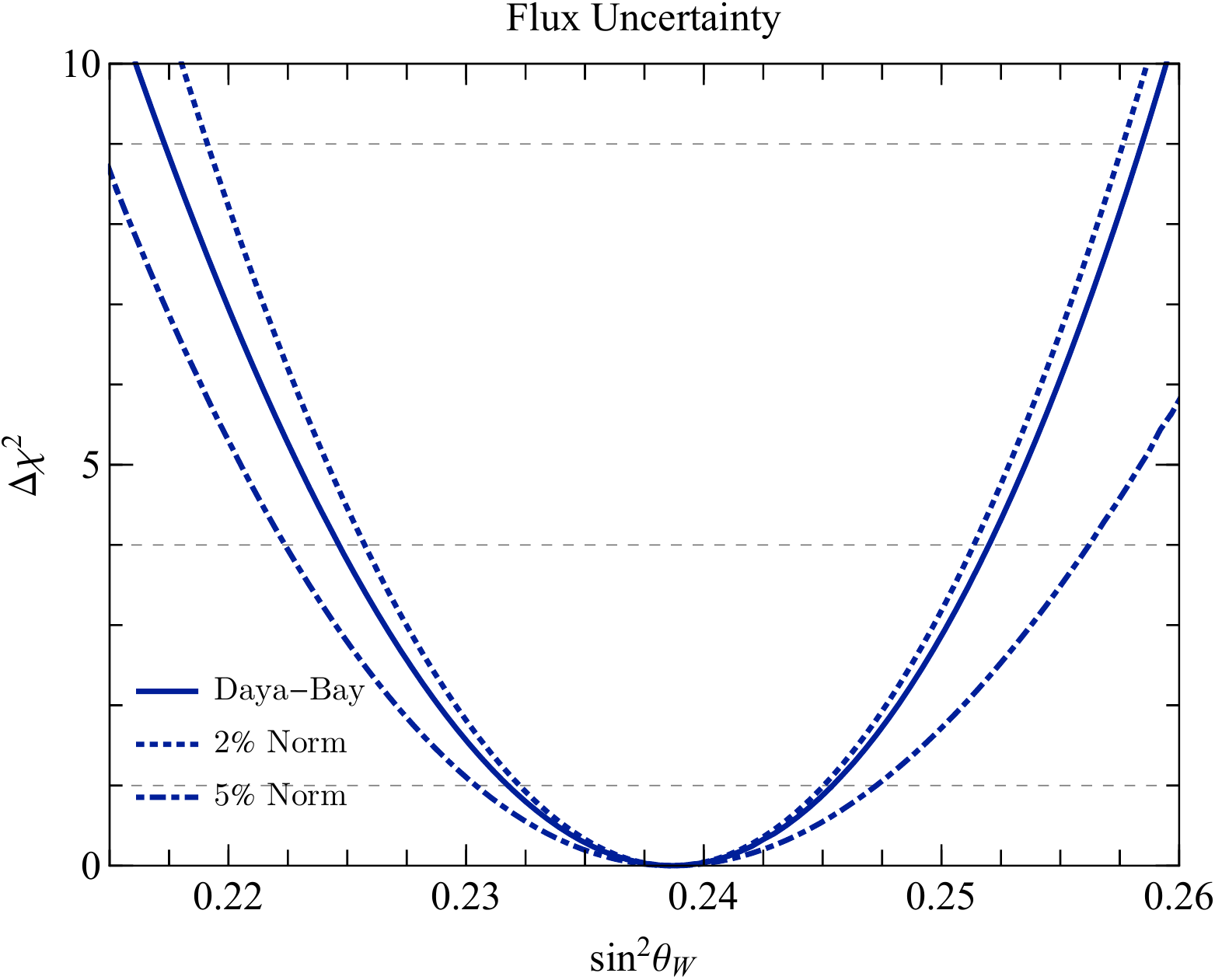}
\caption{Precision on the determination of $\sin^2\theta_W$ for different flux uncertainties: Daya Bay covariance matrix~\cite{Adey:2019ywk} (solid blue), 2\% theory normalization error~\cite{Mention:2011rk, Huber:2011wv} (dotted blue) and 5\% flux normalization error~\cite{Hayes:2013wra, Hayes:2016qnu, Sonzogni:2017wxy} (dashed blue).}
\label{fig:flux-uncertainty}
\end{figure}

As mentioned above, in view of the recent reactor antineutrino anomaly, one may find more robust to do precision analysis with the Daya Bay flux determination.
Nevertheless, it is a useful exercise to see the dependence of the sensitivity to $\sin^2\theta_W$ on the reactor antineutrino flux uncertainty.
To do so, we perform our last analysis in this manuscript under three assumptions regarding the flux uncertainties, in Fig.~\ref{fig:flux-uncertainty}: Daya Bay covariance matrix~\cite{Adey:2019ywk} (solid blue), 2\% theory normalization error~\cite{Mention:2011rk, Huber:2011wv} (dotted blue) and 5\% flux normalization error~\cite{Hayes:2013wra, Hayes:2016qnu, Sonzogni:2017wxy} (dashed blue).
As we can see, the Daya Bay flux determination is already quite good for our benchmark sensitivity. 
By reducing the flux uncertainty from Daya Bay's covariance matrix to a 2\% overall normalization, the sensitivity improves marginally. 
We therefore conclude that, although an improved flux model or determination would be beneficial, this is not a bottleneck in our experimental proposal.

\section{Conclusions}
\label{sec:conclusions}
In this paper, we have analyzed the sensitivity to the weak mixing angle of an experimental configuration like the recently proposed vIOLETA experiment, a Skipper CCD detector deployed 12 meters away from the core of the commercial nuclear reactor Atucha~II in the province of Buenos Aires, Argentina.
We have analyzed the impact of several experimental aspects on the estimated sensitivity: quenching factor and its uncertainty, background rate and spectral shape, and total exposure.
We have quantified the crucial role of the quenching factor on the sensitivity.
Nevertheless, the role of systematic uncertainties in the quenching depends on the quenching itself.
For the Lindhard quenching model, these systematics play a small role, while the opposite happens for the conservative Chavarria parametrization.
As expected, the background rate is also critical to the success of a neutrino experiment leveraging the Skipper CCD technology.
A flat background of 1~kdru would allow for a competitive measurement of the weak mixing angle, while a rate of 10~kdru would be prohibitive.
Moreover, if the background follows a $E_I^{-1}$ spectral shape, similar to the signal, mitigation becomes of the utmost importance.
Finally, our findings also show that the experimental sensitivity is not systematics limited for a 3~kg-year exposure. 
Therefore, better measurements may be achieved by deploying a larger detector.
Our findings show that, under realistic assumptions, the measurement of the weak mixing angle performed by this experimental setup would be competitive with other experiments, including DUNE.
This measurement would be one of the very few competitive determinations of the weak mixing angle using neutrinos, which is particularly important given the unsettled NuTeV discrepancy.
We hope our results can be useful to experimentalists as a prioritization guide for maximizing the physics output of a reactor neutrino Skipper CCD experiment.


\section*{Acknowledgements}
We thank Ivan Sidelnik for invaluable comments on the manuscript and Ricardo Piegaia for useful discussions.
SRA would like to thank Fermilab's theory group for the kind hospitality during the first stages of this work.
Fermilab is operated by the Fermi Research Alliance, LLC under contract No. DE-AC02-07CH11359 with the United States Department of Energy. 
SRA acknowledges the support of the Spanish Agencia Estatal de Investigacion and the EU ``Fondo Europeo de Desarrollo Regional'' (FEDER) through the projects PID2019-108892RB-I00/AEI/10.13039/501100011033 and FPA2016-78645-P; the ``IFT Centro de Excelencia Severo Ochoa SEV-2016-0597''; as well as the European Union's Horizon 2020 research and innovation programme under the Marie Sklodowska-Curie grant agreement 690575-InvisiblesPlus. 

\bibliographystyle{kpmod}
\bibliography{CEvNS.bib}
\end{document}